\documentclass[a4paper,11pt]{article} 
\usepackage{authblk}
\usepackage{fullpage}
\usepackage{graphicx}
\usepackage{xcolor}
\usepackage{algorithm}
\usepackage{algpseudocode}
\usepackage{amssymb}
\usepackage{amsthm}
\usepackage{amsmath}
\usepackage{mathrsfs}
\usepackage[mathcal]{euscript}
\usepackage{bm}
\usepackage{subcaption}
\usepackage{mathabx}
\usepackage[square,numbers,comma,sort&compress]{natbib} 
\usepackage[colorlinks,urlcolor=cobalt,citecolor=cobalt,linkcolor=cobalt,pdftex, pdfauthor = Lukas Exl]{hyperref}
\usepackage{tabularx}
\usepackage{setspace} 
\usepackage{wasysym}

\definecolor{airforce}{rgb}{0.16,0.32,0.75}
\definecolor{cobalt}{rgb}{0.0,0.28,0.67}

\definecolor{azu}{rgb}{0.0,0.4,1.0}

\title{\Large{\textbf{Magnetic microstructure machine learning analysis}}} 
\author[1,2,3]{Lukas Exl}
\affil[1]{\small WPI c/o Faculty of Mathematics, University of Vienna, 1090 Vienna, Austria.}
\affil[2]{\small Institute of Mathematics, University of Vienna, 1090 Vienna, Austria.} 
\affil[3]{\small Institute for Analysis and Scientific Computing, Vienna UT, 1040 Vienna, Austria.} 
\author[4]{Johann Fischbacher}
\author[4]{Alexander Kovacs}
\author[4]{Harald Oezelt}
\author[4]{Markus Gusenbauer}
\author[4,5,6]{Kazuya Yokota}
\author[5,6]{Tetsuya Shoji}
\author[7]{Gino Hrkac}
\author[4]{Thomas Schrefl\thanks{\texttt{tschrefl@gmail.com}}}
\affil[4]{Department for Integrated Sensor Systems, Danube University Krems, Viktor Kaplan Str. 2/E, 2700 Wiener Neustadt, Austria}
\affil[5]{Toyota Motor Corporation, 1 Toyota-cho, Toyota, Aichi 471-8572, Japan}
\affil[6]{Technology Research Association of Magnetic Materials for High-efficiency Motors (MagHEM), Higashifuji-Branch, 1200 Mishuku, Susono Shizuoka 410-1193, Japan}
\affil[7]{College of Engineering, Mathematics and Physical Sciences, The University of Exeter, Exeter, EX4 4SB, United Kingdom}
\begin{document}

\date{}
\maketitle
\noindent\textbf{Abstract.}
We use a machine learning approach to identify the importance of microstructure characteristics in causing magnetization reversal in ideally structured large-grained Nd$_2$Fe$_{14}$B permanent magnets. 
		The \textit{embedded Stoner-Wohlfarth method} is used as a reduced order model for determining local switching field maps which guide the data-driven learning procedure. The predictor model is a random forest classifier 
		which we validate by comparing with full micromagnetic simulations in the case of small granular test structures. In the course of 
		the machine learning microstructure analysis the most important features explaining magnetization reversal were found to be the misorientation and the position of the grain within the magnet.  
		The lowest switching fields occur near the top and  bottom edges of the magnet. While the dependence of the local switching field on the grain orientation is known from theory, 
		the influence of the position of the grain on the local coercive field strength is less obvious. 
		As a direct result of our findings of the machine learning analysis we show that edge hardening via Dy-diffusion leads to higher coercive fields. \\ 

\noindent\textbf{Keywords.} permanent magnets, machine learning, (data-driven) model order reduction, embedded Stoner-Wohlfarth model, feature selection \\

\section{Introduction}

Permanent magnets are widely used in modern society. The high performance magnet market is dominated by Nd$_2$Fe$_{14}$B magnets. 
The six major application areas are acoustic transducers, air conditioning, electric bikes, wind turbines, hybrid and electric cars, and hard disk drives \cite{Constantinides2016}. 
Growing demands for permanent magnets are predicted for green technology applications such as sustainable energy production and eco-friendly transport.  
The generator of a direct drive wind mill requires high performance magnets of 400 kg/MW power; and on average a hybrid and electric vehicle needs 1.25 kg of high end permanent magnets \cite{yang2017ree}.  
Another rapidly growing market is electric bikes. The global demand for rare earth elements in permanent magnets will exceed 50 thousand tons per year in 2025 \cite{yang2017ree}.  
With the quest for rare-earth reduced or rare-earth free permanent magnets \cite{skokov2018heavy}, an optimal control of the magnet's microstructure becomes increasingly important. 
In other fields of materials research, data driven machine learning approaches have been applied recently, in order to obtain a deeper understanding of the material's microstructure on its properties. 
Mangal and Holm \cite{mangal2018applied2} combined crystal plasticity based simulation with machine learning techniques for predicting stress hot-spots in polycrystalline metals. 
Using random-forest based machine learning they correlate the formation of grains with high stress by uniaxial tensile deformation with local microstructural features that describe crystallography, geometry, and connectivity. 
In another paper \cite{mangal2018comparative}, they addressed the problem of feature selection for the classification of stress hot spots. They showed that a proper set of microstructural features is required, in order to find out what microstructural characteristics will cause high local stress during tensile deformation.

Modern Nd$_2$Fe$_{14}$B permanent magnets show a granular structure. Ideally, the grains are separated by a nonmagnetic grain boundary phase \cite{fidler1989electron}. 
In order to improve the isolation  of the grains by a nonmagnetic Nd-rich grain boundary phase, a high Nd content and a dopand such as Al \cite{fidler1989electron} or Ga \cite{sasaki2016formation} are required. In this work we investigate the influence of the microstructure on the local coercivity of permanent magnets with ideal structure. We assume grains that are completely separated by a nonmagnetic phase, and we do not introduce any soft magnetic defects. Using machine learning techniques we identify the microstructural characteristics that may cause weak grains, which are defined as the grains that will reverse first when an increasing opposite field is applied to the magnet. By neglecting defects and ferromagnetic grain boundaries we focus on the effects of key structural features that are common to any polycrystalline material such as grain size, grain shape, grain sphericity, and crystallographic orientation.

\section{Methods}
\subsection{Dataset generation}

\begin{figure} 
\centering
\includegraphics[scale=0.48]{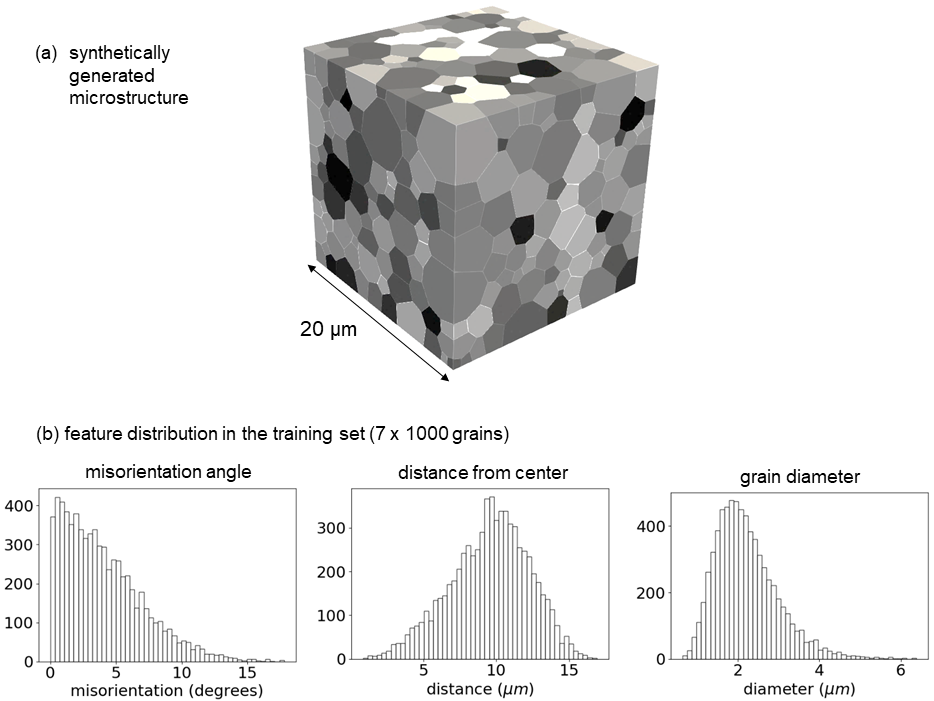}
\caption{\label{fig:model} (a) Example of a synthetically generated grain structure. (b) Distributions of features of grains in the training set: Misorientation angle (for the scenario with a standard deviation of the misorientation angle of 5 degrees), distance of the grain from the magnet's center, and grain diameter.}
\end{figure}

We investigate magnetic multigrain structures in view of their switching field distribution aiming at predicting grains with low switching field (weak grains) and those with high switching field (strong grains), respectively. 
We generate synthetic microstructures consisting of polyhedral grains using the software \texttt{Neper} \cite{quey2011large,quey2018optimal}. 
We use the default grain growth parameter which gives a wider grain size distribution and higher grain sphericities than a standard Voronoi tessellation. The grain size normalized by the average grain size, $D/\langle D \rangle$, follows a lognormal distribution with a standard deviation of 0.35. The sphericity $s$ is a metric for the shape of the grains \cite{quey2018optimal}. It is defined as the ratio of the surface area of a sphere with equivalent volume to the surface area of the grain. The quantity $1-s$ follows a lognormal distribution with a mean of 0.145 and a standard deviation of 0.03. We investigate three scenarios depending on the standard deviation of the misorientation angle of the anisotropy direction: $\sigma_\theta = 0^\circ$, $\sigma_\theta = 5^\circ$, and $\sigma_\theta = 15^\circ$.  For each scenario 10 synthetic microstructures with 1000 grains each were generated. Seven structures were randomly selected to form the training set. The remaining three structures build the test set.
Fig.~\ref{fig:model}a shows a typical microstructure. Fig.~\ref{fig:model}b shows the distributions of the some features in the training set: The misorientation angle of the anisotropy axes, the distance of the grain from the magnet's center, and the grain size. The training set contains $7\times1000$ grains.  

\begin{figure}
\centering
\includegraphics[scale=0.45]{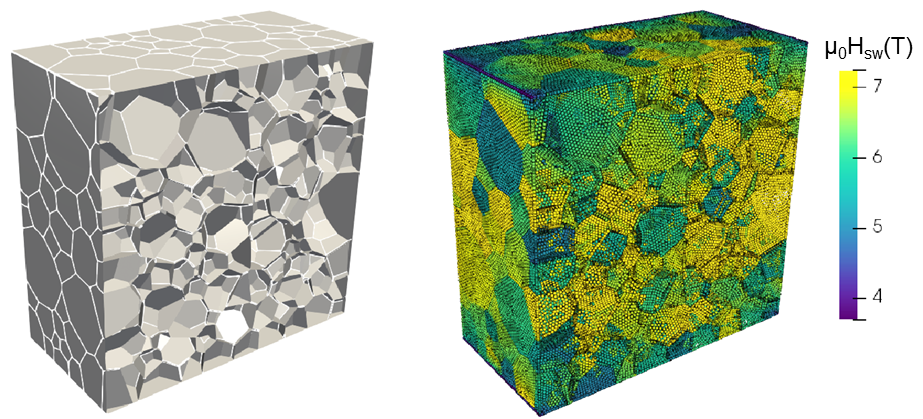}
\caption{\label{fig:multigrain} Cut through a synthetic microstructure to visualize the grain shapes (left) and the local switching field at evaluation points (right). }
\end{figure}

Switching field values are calculated near the surface of the grains which serve as underlying datasets for the \textit{microstructure machine learning analysis}. Fig.~\ref{fig:multigrain} shows a cut through the grain structure,  
the locations of the field-evaluation points, and the calculated switching fields.
Since there are no pinning sites for domain walls within a grain a reversed domain will expand through the grain once it is nucleated. Therefore, the minimum value of the switching fields within a grain defines its reversal field which is used for machine learning.  For the simulations we use the material properties of Nd$_2$Fe$_{14}$B (anisotropy constant $K_1 = 4.9$~MJ/m$^3$, spontaneous magnetic polarization $\mu_0 M_\mathrm{s} = 1.61$~T, and exchange constant $A = 8$~pJ/m \cite{coey2010}) and a mean grain size of 2 $\mu$m. Here $\mu_0$ is the permeability of vacuum.

\subsubsection{Embedded Stoner-Wohlfarth method}

\begin{figure}
\centering
\includegraphics[scale=0.14]{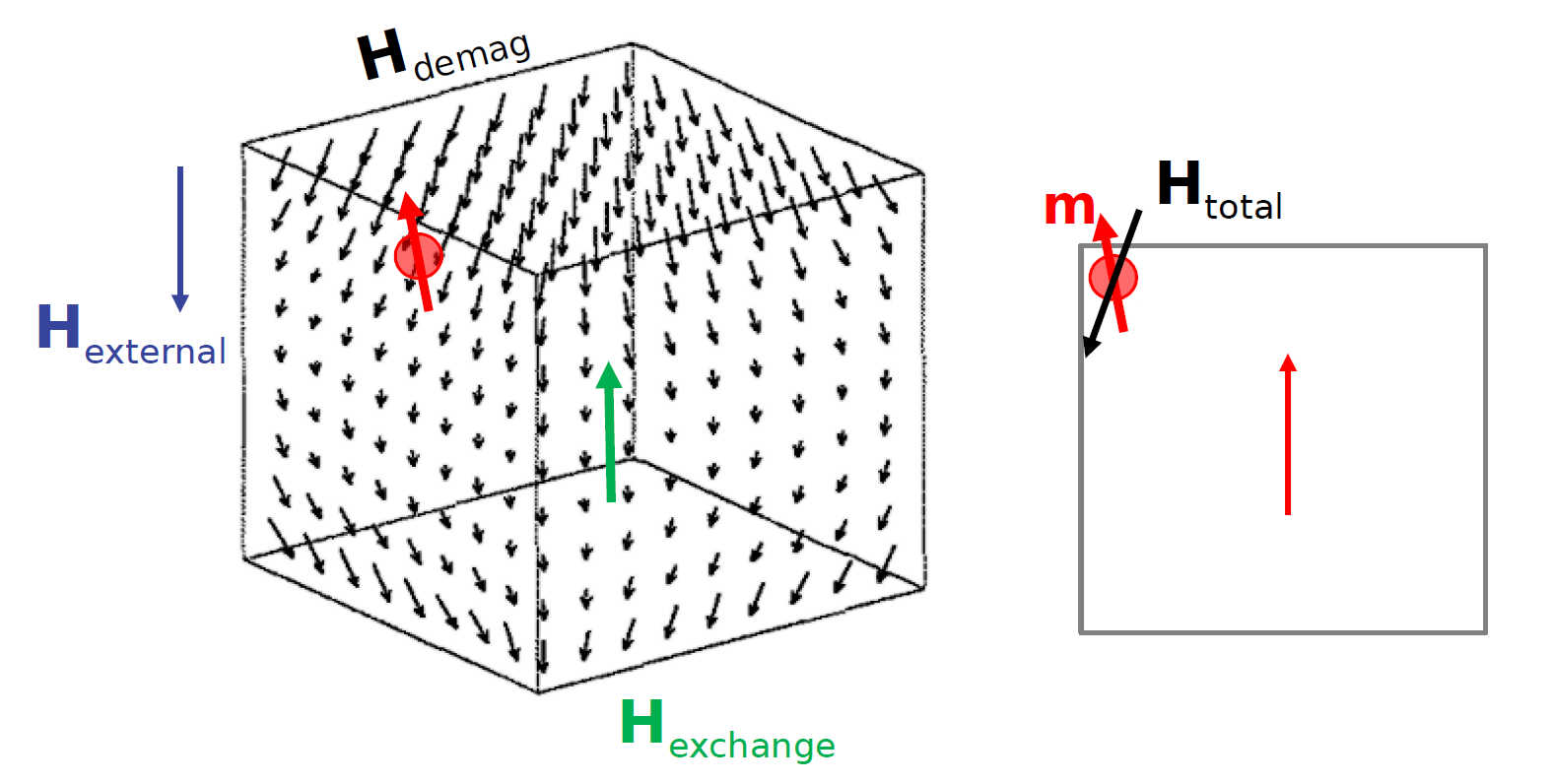}
\caption{\label{fig:esw} Field components in the embedded Stoner-Wohlfarth method (left) and total field (right).}
\end{figure}

The micromagnetic calculation of switching fields in permanent magnet models relies on hysteresis computation usually using numerous successive total energy minimization steps for slightly varying external field strength. 
This is only feasible for models in the nanometer regime with a few grains. Since our data driven approach requires hundreds of grains our models are too large for conventional micromagnetic simulations. Hence we apply a reduced order model for the prediction of critical fields, 
called the \textit{Embedded Stoner-Wohlfarth method (ESW)} \cite{fischbacher2017searching}. 
The approach has its origin in the work of Schrefl and Fidler \cite{schrefl1992numerical} and adapts the original Stoner-Wohlfarth model for small ferromagnetic particles in a way to additionally account for long-range interactions of uniformly magnetized grains. 
To this end the stray field computations are accomplished by analytical formulas for polyhedral geometries \cite{guptasarma1998new}.  First the total field is calculated 
\begin{align}\label{total_field}
 \mathbf{h}_{\textrm{tot}} = \mathbf{h}_{\textrm{ext}} + \mathbf{h}_{\textrm{demag}} + \mathbf{h}_{\textrm{x}},
\end{align}
the sum of external, demagnetizing and exchange field. The perpendicular component of the demagnetizing field grows with no bound towards the edges of a polyhedron, which is compensated by the exchange field \cite{rave1998corners}. 
We define the parallel (to the magnetization) component of the exchange field as $h_{\textrm{x}} = (1/(\mu_0 M_\mathrm{s}))A/d^2$ \cite{fischbacher2017searching} and set the perpendicular component to zero. Fig.~\ref{fig:esw} shows the field components in a cubic particle. The distance $d$ is $1.2 L_\mathrm{ex}$. The exchange length $L_\mathrm{ex}$ is $\sqrt{A/(\mu_0 M_\mathrm{s}^2)}$.
According to Stoner-Wohlfarth the switching field \cite{stoner1948mechanism} of a small uniformly magnetized particle can be given in terms of the angle $\psi$ between the easy axis 
and the total field by the formula \cite{kronmuller1987angular}
\begin{align}\label{sw_classic}
h_{\textrm{sw}} = f(\psi)\,h_N,\, f(\psi) = (\sin^{2/3}\psi + \cos^{2/3}\psi)^{-3/2},
\end{align}
where $h_\textrm{N}$ is the ideal nucleation field \cite{kronmuller2007general},  $h_\textrm{N} = 2\,K_1/(\mu_0\,M_\textrm{s})$. In a hard magnetic particle the easy axis coincides with the magneto-crystalline anisotropy direction.  
The Stoner-Wohlfarth switching field (\ref{sw_classic}) is evaluated  for varying external field locally at target points a distance $d$ away from 
the surface of the polyhedral grains \cite{bance2014grain,fischbacher2017searching}, where the angle between the anisotropy direction and the total field \eqref{total_field} is taken. Please note that in the remanent state the magnetization can be assumed to be approximately parallel to the anisotropy direction. The local switching field at a target point is the smallest value of $|\mathbf{h}_{\textrm{ext}}|$ which makes the total field greater than the value obtained from \eqref{sw_classic}, that is $|\mathbf{h}_{\textrm{tot}}| > h_{\textrm{sw}}$. Then we compute the minimum switching field over all target points of a grain. This minimum value is the switching field of the grain, which is then used for labeling weak and strong grains in the subsequent machine learning task.  

\subsubsection{Microstructure attributes}
Our main intuition is that weak points in permanent magnet grain structures can be well understood by their (mainly) geometrical microstructure attributes.  
The machine learning approach will assign these features to each grain together with the grain label (weak or strong grain) according to calculated switching field values using embedded Stoner-Wohlfarth as an effective reduced model. 
The following geometrical attributes are assigned:
\begin{itemize}
 \item The z-coordinate of the center of the polyhedral grain (\texttt{z-position}),
 \item The distance to the center of the magnet (\texttt{distance}),
 \item The diameter of the polyhedron (\texttt{diameter}) defined as the diameter of a sphere with equivalent volume,
 \item The number of next neighbor grains (\texttt{no of neighbors}),
 \item The sphericity of the grain (\texttt{sphericity}),
 \item The absolute deviation of the current grain diameter and the average diameter of the next neighbors (\texttt{diam variation}),
 \item The maximum dihedral angle of the polyhedron (\texttt{max dihedral angle}),
 \item The minimum dihedral angle of the polyhedron (\texttt{min dihedral angle}).
\end{itemize}

\begin{figure}
\centering
\includegraphics[scale=0.36]{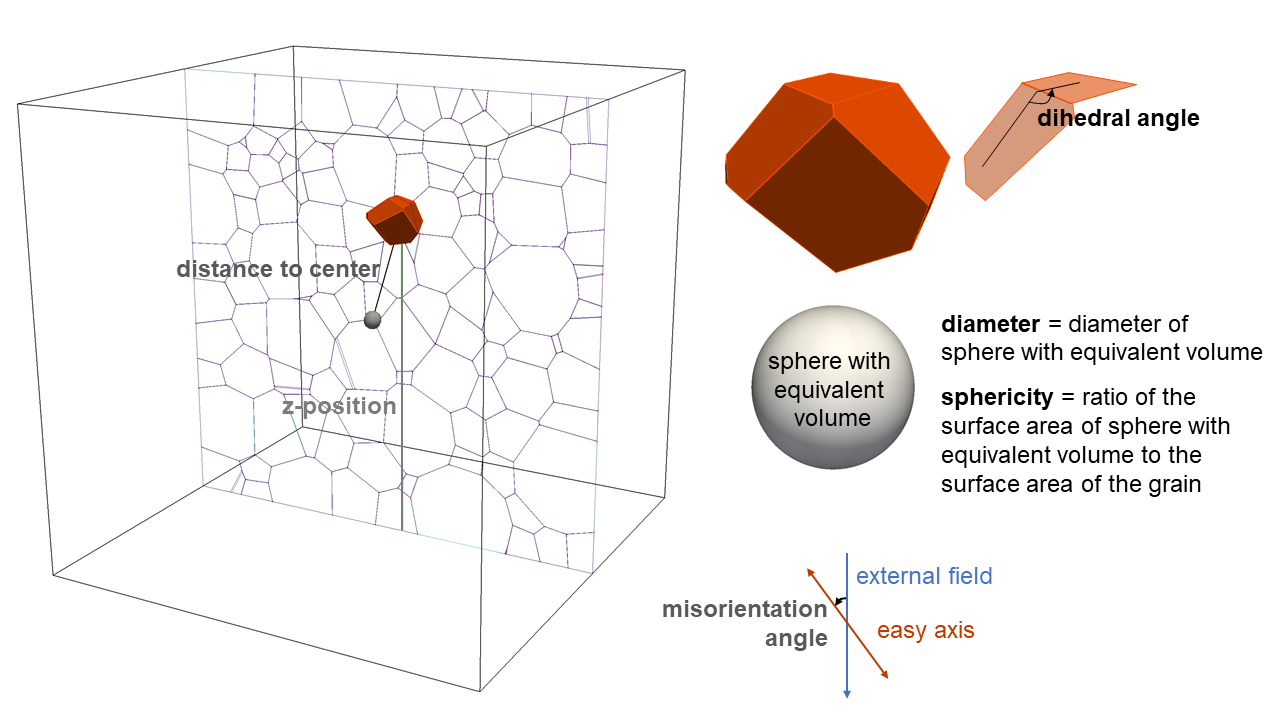}
\caption{\label{fig:features} Sketch of the selected descriptors: Distance to center, z-position, misorientation, diameter, sphericity and dihedral angle, where we use the maximum and minimum dihedral angle of a grain.}
\end{figure}

In permanent magnets the magnetocrystalline anisotropy energy is expressed by $K_1\sin^2 (\varphi - \theta)$
where $\varphi$ is the angle between the magnetization and the saturation direction and $\theta$ the angle between the $z$-axis of the tetragonal crystal and the saturation direction. In the embedded Stoner Wohlfarth model the orientation dependence of the switching field expressed by \eqref{sw_classic}  describes the reduction of the anisotropy field by a factor that depends on the angle $\psi$ between the easy axis and the total field \eqref{total_field}. 
Hence, additionally to geometrical features we assign the orientation of the easy axis for
each grain.
\begin{itemize}
 \item The orientation angle $\theta$ of the grains (\texttt{misorientation}).
\end{itemize}
Fig.~\ref{fig:features} shows a sketch of some of the descriptors. 

\begin{figure}
\centering
\includegraphics[scale=0.5]{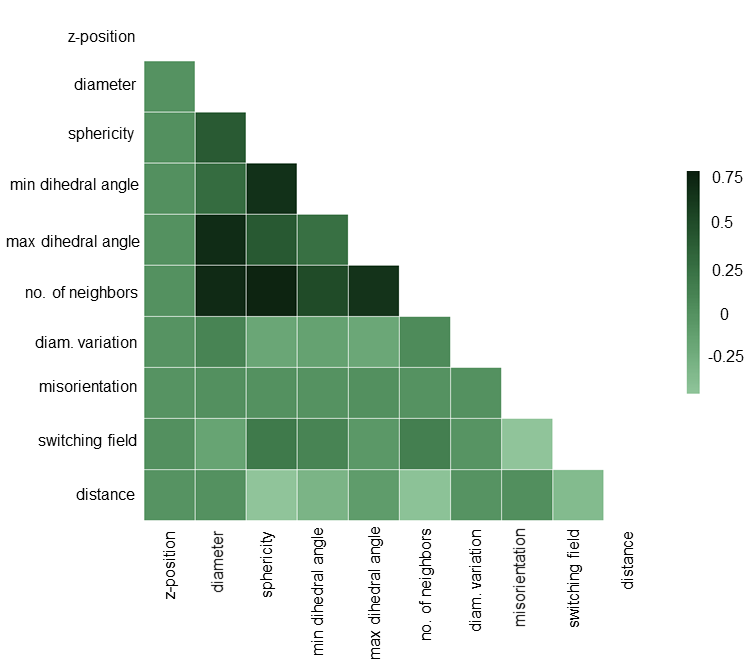}
\caption{\label{fig:corr} Correlation matrix of the selected descriptors including the local switching field. All correlation coefficients are smaller or equal 0.76.}
\end{figure}

The contribution of each of the above attributes in predicting weak and strong grains is studied statistically by the machine learning approach.
These features represent an already preselected and rather uncorrelated subset of a larger possible set of attributes. 
For instance, attributes like the surface area, the volume and the diameter of the grains exhibit correlation coefficients above $0.95$. Pearson's correlation coefficient \cite{lee1988thirteen} is a measure of the tendency of the features to increase or decrease together. Therefore, we only took one ``representative'' when the correlation coefficient between a pair of features was greater than $0.76$. For example, we only take the grain diameter and drop surface area and volume. The correlation matrix for the selected descriptors is shown in Fig.~\ref{fig:corr} which in addition includes the local switching field attribute. 
Remarkably, the z-position is initially uncorrelated with the local switching field but gets a decisive role in explaining weak and strong grains as indicated by its feature importance (compare with Sec.~\ref{sec:m3la}). 

\subsection{Machine learning methods}
\begin{figure}
\centering
\includegraphics[scale=0.4]{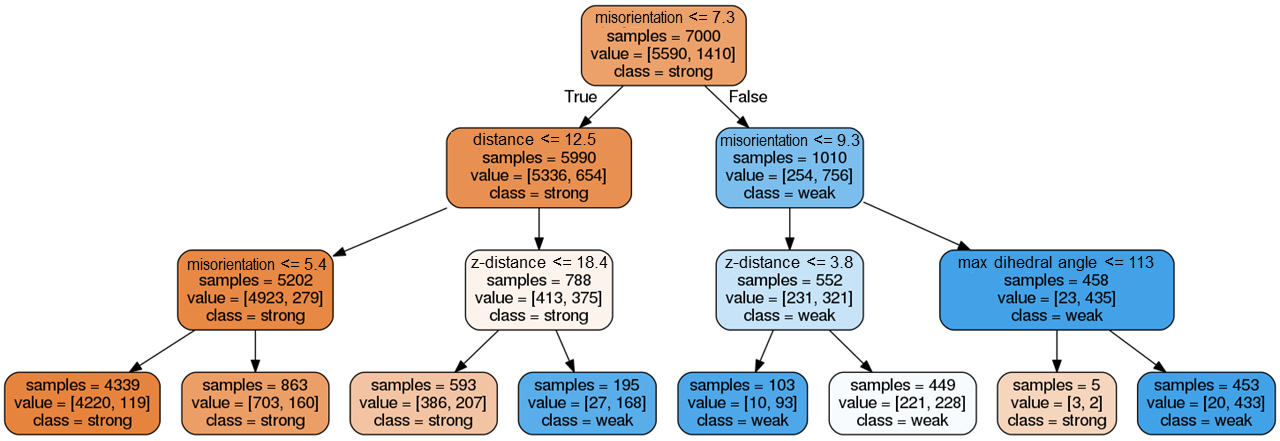}
\caption{\label{fig:dtree} Example of individual decision tree. First, the training set consisting of 7000 grains is split into 5990 strong grains and 1010 weak grains depending on the crystallographic orientation. The two nodes in the second level are split depending on the distance and on the orientation, respectively. In the third level the orientation, the z-position, and the maximum dihedral angle become decisive features.}
\end{figure}

Machine learning is a statistical approach that aims at automating analytical model fitting for data analysis, for instance finding clusters/structures in data or generating data-based predictive decision tools. 
For a very comprehensive introduction to machine learning the reader is referred to \cite{geron2017hands}. 
We use so-called \textit{supervised learning}, where the training data also includes the true solutions. In our case, the training data consist of grains together with their predictors (the geometrical microstructure features or descriptors) 
and labels (their switching fields). We aim at classifying weak grains, that is, predicting those feature classes which exhibit a switching field below a certain threshold (class A) and above it (class B), respectively. Beside \textit{classification} 
a second common supervised learning task is \textit{regression}, which would try to predict values instead of classes. However, similar as in \cite{mangal2018applied2} we decide to use a \textit{random-forest} (RF) algorithm to not 
only built up a predictive binary classifier but also get insight into the \textit{feature importance} causing weak grains \cite{breiman2001random}. Random-forest algorithms are \textit{bagging methods} built up by combining predictions of individual \textit{decision trees} 
trained over randomly generated sub-training samples. At any instance an average of the individual estimators is taken to generate the ensemble model. An example of one decision tree with depth three is given in Fig.~\ref{fig:dtree}.  

An important and non-trivial task is the performance measure of a classifier. 
The \textit{accuracy} of a model is the amount of correctly predicted instances relative to all instances. 
Depending on the tightness of the threshold of switching field value (= decision threshold) used for classifying weak grains any accuracy could be achieved. For instance, if the smallest $10\%$ of all grains are labeled as weak a classifier which invariably 
predicts strong grains will have a $90\%$ accuracy. A way out is to determine the \textit{confusion matrix} of a binary classifier, 
that is to count the number of times instances of one class (strong or weak grain) are classified correctly (true weak or strong) or incorrectly (false weak or strong), respectively. 
The ratio of the number of true weak grains and all grains classified as weak is called \textit{precision}. 
A high precision means that few strong grains are erroneously classified as weak, where possibly many weak instances can still be erroneously classified as strong.  
Instead, the so-called \textit{recall} is the ratio of the number of true weak grains and the sum of true weak and false strong instances. 
A high recall means that few weak grains are erroneously classified as strong, where possibly many strong instances can still be erroneously classified as weak. 
Obviously there is a trade-off between precision and recall. The harmonic mean of precision and recall is the \textit{f1-score} 
of the binary classifier. A random-forest model depends on various hyperparameters. It is good practice to optimize the hyperparameters according to the problem. We maximized the \textit{f1-score} by searching optimal values for the tree depth, the number of trees, and the number of features to consider when splitting a node. We calculated the confusion matrix with respect to the test set. Another performance measure is the \textit{receiver operating characteristic} (ROC) curve which plots the true positive rate versus 
the false positive rate. The area under the ROC curve (AUC) is a common evaluation metric whereas values close to $1$ indicate a good classifier.

In a decision tree important features are likely to appear closer to the root of the tree, whereas unimportant features are found near the leaves  or not at all. Estimates of the feature importance in a random-forest classifier can be calculated by the average depth at which it appears across all trees. 
Another approach to determine feature importance is a model-agnostic version called \textit{model reliance}, where feature importance is indicated by the amount of increase of 
model error, for example measured by the $\textrm{AUC}$, by fitting a model after permuting the features \cite{molnar2018,fisher2018model}.

\section{Results}
\subsection{Micromagnetic validation}
\begin{figure}
\centering
\includegraphics[scale=0.30]{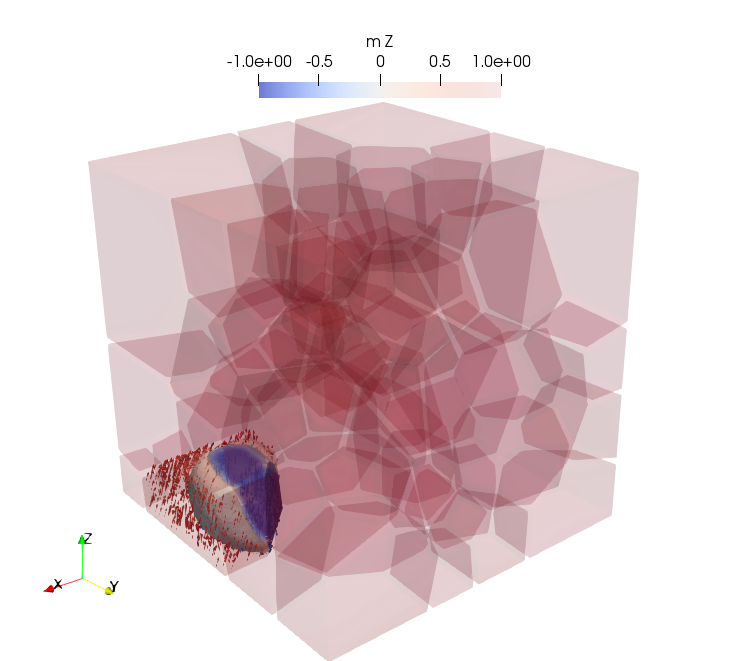}
\caption{\label{fig:mm} Example of identification of the weakest grain with the full micromagnetic simulation for one test structure. Magnetization reversal starts in the highlighted grain on the bottom-left where a reversed domain is already formed.}
\end{figure}

In the case of structures consisting of very few grains we can validate our approach with full micromagnetic computation including the conventional determination of the magnetostatic field via Maxwell's equations. 
The question is whether a trained random-forest model can predict where magnetization reversal will start. We create 100 granular structures consisting of only 64 grains each with a mean grain size of 50~nm.
We split the data structures into 80 training structures and 20 test structures. For each structure the grain orientations in z-direction are set randomly according to a uniform distribution for the azimuthal angle and a zero-mean normal distribution with the standard deviation of $5$ degrees. We first label the grains as "weak" or "strong" according to the switching fields computed by the embedded Stoner-Wohlfarth model. Then we train a random-forest model on the training set using the Python library \texttt{Scikit-Learn} \cite{pedregosa2011scikit}. In order to validate the model, we perform full micromagnetic simulations using the finite element method \cite{fischbacher2018micromagnetics}. Following the demagnetizaton curve we compute the grain and corresponding switching fields where magnetization reversal starts. This identifies the true weakest grains in the test set (see Fig.~\ref{fig:mm}). In 16 out of the 20 test cases the random-forest prediction of the weakest grain coincides with the results from full micromagnetic simulations. 

We also estimated the model error of the embedded Stoner-Wohlfarth model. In 18 out of 20 cases the weakest grains according to the embedded Stoner-Wohlfarth model and full micromagnetic switching fields coincide. This discrepancy reflects the model error mainly corresponding to the simplified stray flied calculation in the embedded Stoner-Wohlfarth model which does not take into account reversal magnetization rotations before switching.

Considering both, the model error of the embedded Stoner-Wohlfarth model and the performance measure of the random-forest model (see Tab.~\ref{tab1}) gives an overall accuracy of 80 percent in accordance with the above validation result.

\subsection{Microstructure machine learning analysis}\label{sec:m3la}
\begin{figure}
\centering
\includegraphics[scale=0.45]{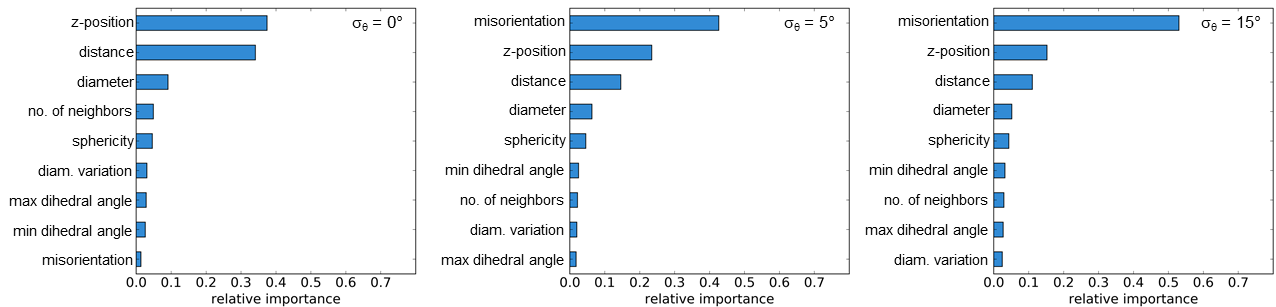}
\caption{\label{fig:fi} Feature importance for the random forest classification in the case of $0,\, 5$, and $15$ degrees standard deviation of the misorientation angle.}
\end{figure}

We use ten multigrain models with $1000$ grains each, where we randomly put aside three models for the validation (this is the test data set).
For the grains in each model we determine the feature values and calculate the true labels by the embedded Stoner-Wohlfarth method in order to supervise the subsequent learning process. The anisotropy directions are set
randomly according to a uniform distribution for the azimuthal angle and a zero-mean normal distribution with a standard deviation of $\sigma_\theta = 0^\circ, \sigma_\theta =5^\circ$, or $\sigma_\theta =15^\circ$ for the polar angle. This determines three different 
scenarios. For each scenario we label grains with a switching field smaller than the 20th percentile of the switching field distribution as "weak" and use the records of the training set to train a random forest model applying the Python library \texttt{Scikit-Learn}\cite{pedregosa2011scikit}.  
Fig.~\ref{fig:fi} shows the feature importance for the three scenarios which was computed using the model agnostic approach \cite{sarkar2018practical} as implemented in \texttt{Skater} \cite{pramit_choudhary_2018_1198885}.

For perfectly aligned grains (0 degree misorientation) there are essentially two most important features, the vertical position of the grain in the magnet ($z$-position) and the distance of the grain from the center of the magnet. When misorientation is introduced, it becomes the most important feature. 
One can clearly observe in Fig.~\ref{fig:fi} that the misorientation becomes more important with higher average misorientation angle. Whereas the dependence of the local switching field on the orientation is expected \cite{kronmuller1987angular}, 
the importance of position of grain within the magnet is less obvious. 

\begin{figure}
\centering
\includegraphics[scale=0.45]{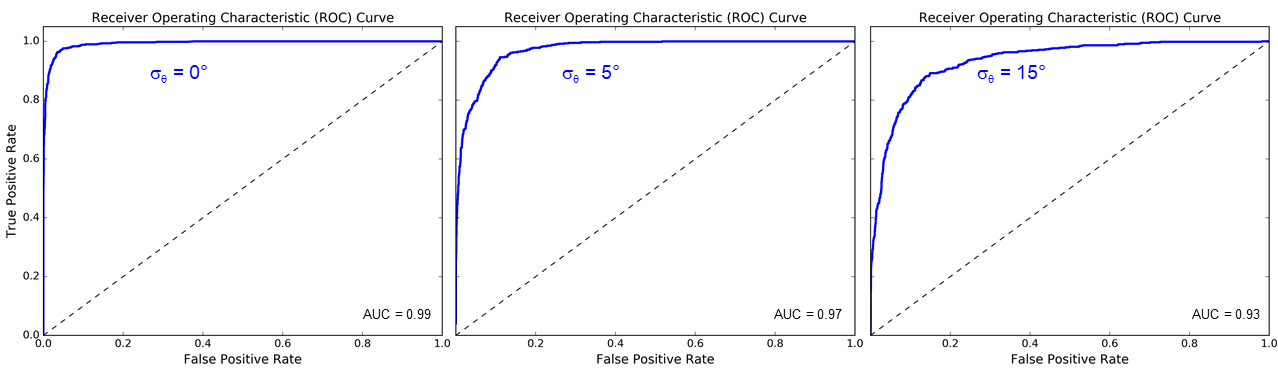}
\caption{\label{fig:roc} Receiver operating characteristic (ROC) curve for the random forest classification in the case of $0,\, 5$, and $15$ degrees standard deviation of the misorientation angle.}
\end{figure}

Tab.~\ref{tab1} shows the confusion matrices and model performance metrics for the random-forest model for $\sigma_\theta = 0^\circ, \sigma_\theta =5^\circ$, and $\sigma_\theta =15^\circ$. Fig.~\ref{fig:roc} shows the ROC curves for the three different scenarios.  The model performance metrics as well as the AUC indicate very high performance of the trained random forest models, whereas a slight decline can be observed with increasing orientation angle.

\begin{table}
\tabcolsep 2pt 
\caption{Confusion matrices for the random forest model for 0, 5, and 15 degrees standard deviation of the misorientation angle. The model performance metrics include accuracy, precision, recall and f1-score.} \label{tab1}
\begin{center}
\begin{tabular}{cc|cc | c c c c}
$\sigma_\theta = 0^\circ$ &  & Predicted & & Performance & metrics & &\\ 
 & & 0 & 1  & accuracy & precison & recall & f1-score\\ \hline
 Actual & 0 & 2369 & 39 & 0.9673 & 0.967 & 0.9673 & 0.9671\\
 & 1 & 59 & 533 & & & &\\ 
\end{tabular}
\end{center}
%

%
\tabcolsep 2pt 
\begin{center}
\begin{tabular}{cc|cc | c c c c}
$\sigma_\theta = 5^\circ$ & & Predicted & & Performance & metrics & &\\ 
 & & 0 & 1  & accuracy & precison & recall & f1-score\\ \hline
 Actual & 0 & 2318 & 91 & 0.924 & 0.9223 & 0.924 & 0.9228\\
 & 1 & 137 & 454 & & & &\\ 
\end{tabular}
\end{center}
%
%

%
\tabcolsep 2pt 
\begin{center}
\begin{tabular}{cc|cc | c c c c}
$\sigma_\theta = 15^\circ$ & & Predicted & & Performance & metrics & &\\ 
 & & 0 & 1  & accuracy & precison & recall & f1-score\\ \hline
 Actual & 0 & 2297 & 112 & 0.896 & 0.8918 & 0.896 & 0.8927\\
 & 1 & 200 & 391 & & & &\\ 
\end{tabular}
\end{center}
\end{table}

\begin{figure}
\centering
\includegraphics[scale=0.42]{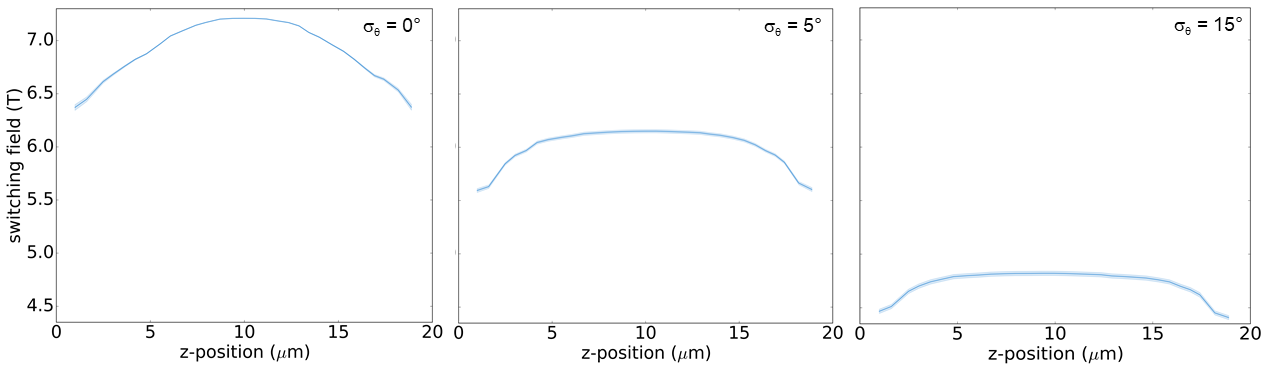}
\caption{\label{fig:lime1} One-way partial dependence based on the vertical position of the grain within the magnet for $0,\, 5$, and $15$ degrees standard deviation of the misorientation angle. A z-position close to $0$ or $20$ $\mu$m indicates a grain near the bottom or top surface of the magnet, respectively.}
\end{figure}

\begin{figure} 
\centering
\includegraphics[scale=0.42]{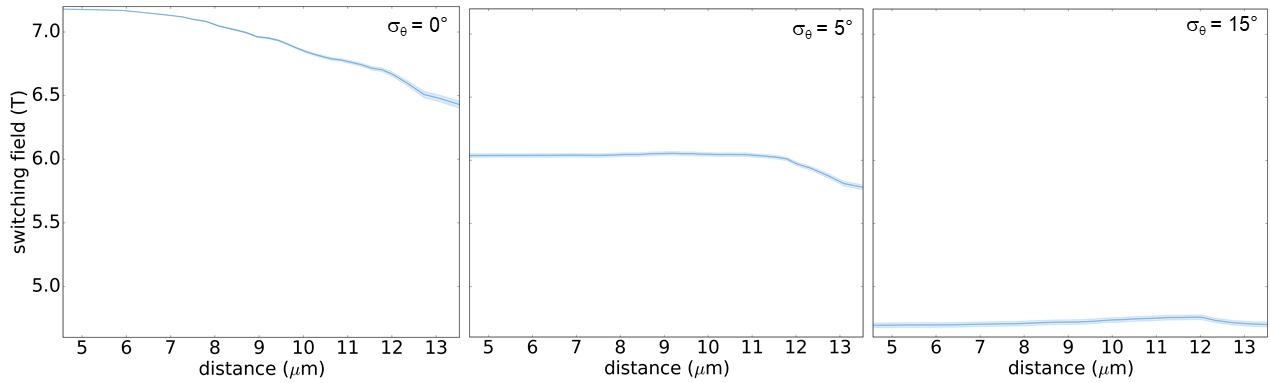}
\caption{\label{fig:lime2} One-way partial dependence based on the distance of the grain to the center of the magnet for $0,\, 5$, and $15$ degrees standard deviation of the misorientation angle.}
\end{figure}

\begin{figure} 
\centering
\includegraphics[scale=0.42]{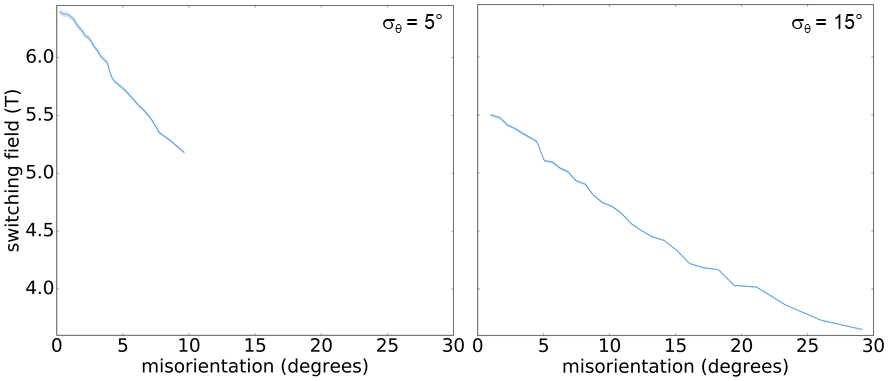}
\caption{\label{fig:lime3} One-way partial dependence based on the misorentation of the anisotropy axes for $5$ (left) and $15$ (right) degrees standard deviation of the misorientation angle.}
\end{figure}

In a second step, we apply random-forest regression to predict the value of the local switching fields of the grains. Then
we can get additional insight into feature dependence by \textit{one-way partial dependence plots} for the random forest predictor using the technique of \textit{local interpretable model agnostic explanation} (LIME) \cite{ribeiro2016should}.  
Fig.~\ref{fig:lime1}, \ref{fig:lime2} and \ref{fig:lime3} show comparisons for different orientation scenarios by one-way dependency based on $z$-position, distance to center and misorientation angle, respectively.  

\section{Discussion}

We applied machine learning techniques in order to correlate the microstructure characteristics with the local magnetization reversal field of large-grained Nd$_2$Fe$_{14}$B permanent magnets. 
In order to focus on general features of polycrystalline materials we assumed an ideal structure: (i) The grains are separated by a nonmagnetic grain boundary phase and (ii) there are no defects with reduced magnetocrystalline anisotropy. 
Though this setting is unrealistic, it can provide clear insight what other features in addition to soft inclusions or ferromagnetic grain boundaries \cite{murakami2014magnetism,zickler2017combined} influence coercivity. 
The data used for machine learning was generated by a reduced order model that makes it possible to treat magnets which are much larger with respect to both grain size and the number of grains than models suitable for micromagnetic simulations. 
For small model size the prediction of the machine learning model can be compared with the results of full micromagnetic simulations. This comparison shows that a random forest classifier can predict the weakest grain in a magnet in 16 out of 20 test cases correctly.

\begin{figure} 
	\centering
	\includegraphics[scale=0.4]{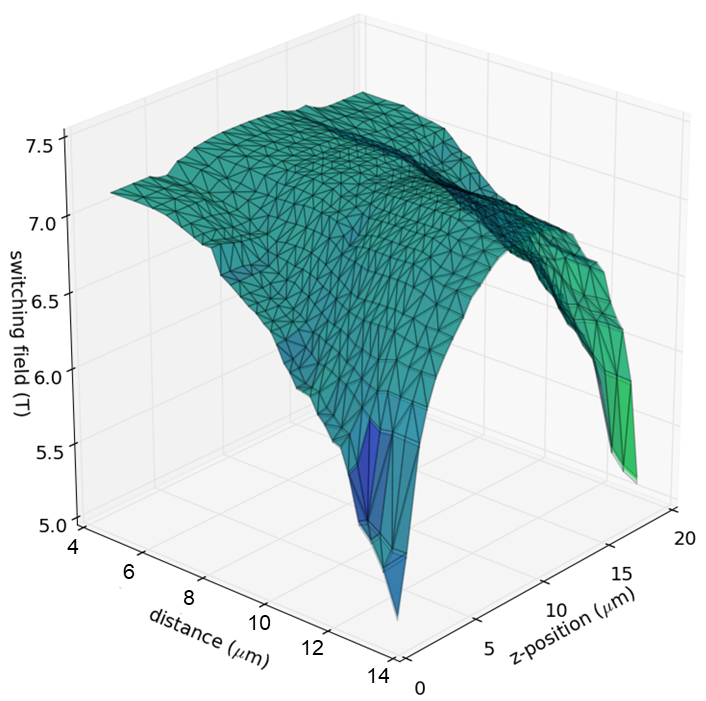}
	\caption{\label{fig:lime4} Two-way partial dependence based on \textit{z-position} and \textit{distance from center} for $5$ degrees standard deviation of the misorientation angle. A z-position close to $0$ or $20$ indicates a grain near the bottom or top surface of the magnet.}
\end{figure}

In order to find out what microstructure features are most significant, we computed the feature importance of a random forest classifier trained with the switching field distribution of 7 polycrystalline samples consisting of 1000 grains each. 
The feature importance was found to depend on the degree of alignment. For a scenario with a standard deviation of the orientation angle of 15 degrees the most important feature is the crystallographic orientation. 
As expected \cite{kronmuller1987angular} the switching field decreases with increasing misorientation angle. The second and third most important features are the vertical position of the grain, and the distance of the magnet from the magnet's center. 
For perfect alignment (zero degree misorientation) these two are the most important features followed by the grain diameter. 
Local interpretable model agnostic explanation \cite{ribeiro2016should} shows that the switching field of a grain is smaller the closer the grain is located to the top or bottom surface of the magnet. 
This dependence is more pronounced for the perfectly aligned grains where the switching field of a grain near the top or bottom is more than 11 percent smaller than that of a grain near the center. 
For the scenario with 15 degrees misalignment the decrease of the switching field based on the  vertical position is 7 percent. 
Similarly, the switching field of a grain decreases with increasing distance from the center of the magnet. 
A two-way partial dependence plot of the switching field as function of \textit{z-position} and \textit{distance from center} shows that the lowest switching fields occur near the top and  bottom edges of the magnet (see Fig.~\ref{fig:lime4}). 
These are the locations where the local demagnetizing field of the magnet reach the highest values \cite{gronefeld1989calculation}. 
Furthermore, near these edges the demagnetizing field is tilted with respect to the magnetization direction which reduces the local Stoner Wohlfarth switching field according to \eqref{sw_classic}.

While the dependence of the local switching field on the grain orientation is known from the basic micromagnetic theory \cite{kronmuller1987angular}, 
the influence of the position of the grain on the local coercive field strength is less obvious. One may argue that strong local demagnetizing field may also occur near the nonmagnetic grain boundary phase inside the magnet 
that may initiate magnetization reversal. The machine learning model shows that this is not the case and the lowest reversal fields always occur near the edges of the magnet. 
These results indicate that local variation of the magnetic properties, which enhances the switching field near the surfaces or edges of the magnet, is sufficient to improve the magnet's performance. 
Possible routes to achieve higher coercive grains locally are grain boundary diffusion \cite{sepehri2013mechanism,sodervznik2016high} and additive manufacturing \cite{huber20163d,li2017additive}. 
Thompson et al. \cite{thompson2017grain} used  electron probe microanalysis to analyze the Dy concentration in diffusion treated sintered magnets and showed that the highest heavy rare-earth concentration occurred 
near the corners of the magnet. A similar local variation of the magnetic properties may be achieved by additive manufacturing. 

\begin{figure} 
	\centering
	\includegraphics[scale=0.36]{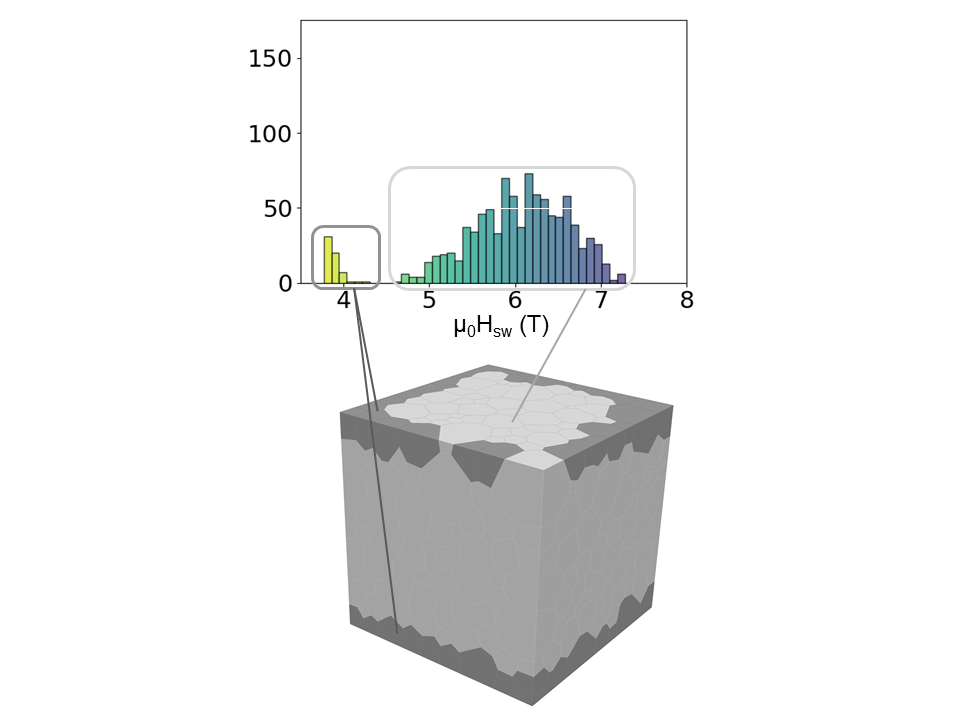}
	\caption{\label{fig:hist_surfhard} Grain structure showing the weakest grains (dark grey) togheter with the switching field distribution of a homogeneous Nd$_2$Fe$_{14}$B model.}
\end{figure}  

\begin{figure} 
	\centering
	\includegraphics[scale=0.36]{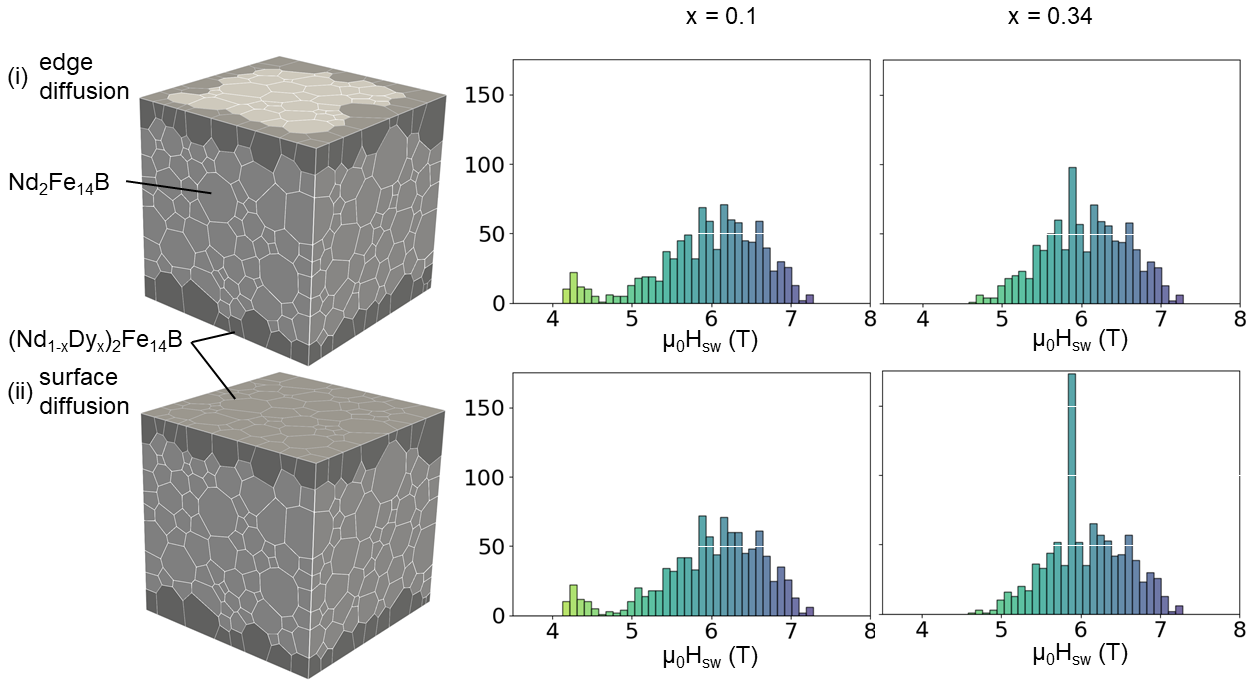}
	\caption{\label{fig:hist_surfhard2} Comparison of switching fields of models with (i) Dy edge diffusion  and (ii) Dy surface diffusion. The grains shown in dark grey refer to Dy containing grains, (Nd$_{0.9}$Dy$_{0.1}$)$_2$Fe$_{14}$B for the histograms on the left hand side and 
		(Nd$_{0.66}$Dy$_{0.34}$)$_2$Fe$_{14}$B for the histograms on the right hand side. The grains shown in light grey are Nd$_2$Fe$_{14}$B.}
\end{figure}

As shown above, machine learning revealed a strong effect the position of the grain within the magnet on the switching field. Indeed,  Fig.~\ref{fig:lime4} shows that the lowest switching fields occur for grains 
located at the edges (near the top and bottom of the magnet and at a large distance of the center). We now take a grain structure from the test set with $5^\circ$ misorientation and analyze its switching field distribution. Fig.~\ref{fig:hist_surfhard} shows the switching field distribution of the grains and the location of the weakest grains. The distribution shows a small peak for $\mu_0 H_{\mathrm{sw}} < 4$~T whereas 
the mean switching field is at 5.9~T and the maximum switching field is at 7.2~T. We can identify the grains with low switching field, which are shown in Fig.~\ref{fig:hist_surfhard}.
As predicted by the machine learning algorithm these are the grains at the top and bottom edges of the magnet.  

In order to show how  Nd$_2$Fe$_{14}$B magnets can be improved by Dy diffusion such as (Nd$_{1-x}$Dy$_{x}$)$_2$Fe$_{14}$B we compare the switching field distribution for different scenarios: 
(i) A sample where the grains near top and bottom edges have higher anisotropy field, and (ii) a sample where the grains near top 
and bottom surfaces have higher anisotropy field. 
Following Oikawa et al. \cite{oikawa2016large} we decrease the spontaneous magnetization $M_s$ linearly with increasing Dy-content. 
For the grains with higher anisotropy field we used (Nd$_{0.9}$Dy$_{0.1}$)$_2$Fe$_{14}$B and (Nd$_{0.66}$Dy$_{0.34}$)$_2$Fe$_{14}$B with a magnetization $\mu_0 M_{\mathrm s} = 1.52$~T and $1.3$~T, respectively.
  When the grains at the top and bottom edges are hardened by Dy diffusion (see Fig.~\ref{fig:hist_surfhard2} (i) ) 
the peak at low fields disappears gradually. The minimum switching field increases from  $\mu_0 H_{\mathrm{sw,min}}  = 3.8$~T without Dy-diffusion to $\mu_0 H_{\mathrm{sw,min}}  = 4.14$~T and $4.62$~T for a Dy content of $x=0.1$ and $x=0.34$ in the grains near the top and bottom edges, respectively. 

The results clearly show that in Dy-free magnets the grains near the top and bottom surface have reduced switching field which in turn reduce the coercive field of the magnet. 
Hardening of the grains near the top and bottom edges by Dy-diffusion avoids these low coercive grains. 
A similar result was achieved by hardening the grains near the top and bottom surface, see Fig.~\ref{fig:hist_surfhard2} (ii). 
This effect may be used in magnet production and may further reduce the heavy rare-earth content while keeping a high coercive field.  

\section{Conclusion}

In summary, we showed that machine learning techniques can be applied to characterize the role of microstructure features in permanent magnets. 
Several application scenarios of machine learning in permanent magnet design can be envisioned ranging from the identification of weak spots to building blocks for the multiscale simulation of hysteresis.

In the example given in this work we identified the location of the weakest grains in ideally structured Nd$_2$Fe$_{14}$B magnets without any defects. The grains with the lowest switching fields are located at the top and bottom edges of the magnet. This suggests that localizing grain boundary diffusion of heavy rare-earth elements to these specific regions only may be sufficient to increase coercivity. Thus, the magnet's performance and temperature stability may be improved with a minimum amount of heavy rare-earth.  

\section*{Acknowledgments}
Financial support by the Austrian Science Fund (FWF) via the project "ROAM" under grant No. 31140 and the SFB "ViCoM" under grant No.41 is acknowledged, as well as support by  
Toyota Motor Corporation, and the future pioneering program Development of Magnetic Material Technology for High-efficiency Motors commissioned by the New Energy and Industrial Technology Development Organization (NEDO).
The computations were partly achieved by using the Vienna Scientific Cluster (VSC) via the funded project No. 71140.

\bibliography{bibref}

\begin{thebibliography}{10}

\bibitem{bance2014grain}
S.~Bance, B.~Seebacher, T.~Schrefl, L.~Exl, M.~Winklhofer, G.~Hrkac,
  G.~Zimanyi, T.~Shoji, M.~Yano, N.~Sakuma, M.~Ito, A.~Kato, and A.~Manabe.
\newblock Grain-size dependent demagnetizing factors in permanent magnets.
\newblock {\em J. Appl. Phys.}, 116(23):233903, Dec. 2014.
\newblock \url{https://doi.org/10.1063/1.4904854}.

\bibitem{breiman2001random}
L.~Breiman.
\newblock Random forests.
\newblock {\em Machine learning}, 45(1):5--32, 2001.

\bibitem{pramit_choudhary_2018_1198885}
P.~Choudhary, A.~Kramer, and datascience.com team.
\newblock {Skater: {Model} Interpretation Library}, Mar. 2018.
\newblock \url{https://doi.org/10.5281/zenodo.1198885}.

\bibitem{coey2010}
J.~M.~D. Coey.
\newblock {\em Magnetism and Magnetic Materials}.
\newblock Cambridge University Press, 2009.

\bibitem{Constantinides2016}
S.~Constantinides.
\newblock Permanent magnets in a changing world market.
\newblock {\em Magnetics Magazine}, Spring 2016:6, 2016.

\bibitem{fidler1989electron}
J.~Fidler and K.~Knoch.
\newblock Electron microscopy of {Nd}-{Fe}-b based magnets.
\newblock {\em J. Magn. Magn. Mater.}, 80(1):48--56, Aug. 1989.
\newblock \url{https://doi.org/10.1016/0304-8853(89)90323-5}.

\bibitem{fischbacher2017searching}
J.~Fischbacher, A.~Kovacs, L.~Exl, J.~Kühnel, E.~Mehofer, H.~Sepehri-Amin,
  T.~Ohkubo, K.~Hono, and T.~Schrefl.
\newblock Searching the weakest link: {Demagnetizing} fields and magnetization
  reversal in permanent magnets.
\newblock {\em Scr. Mater.}, 154:253--258, Sept. 2018.
\newblock \url{https://doi.org/10.1016/j.scriptamat.2017.11.020}.

\bibitem{fischbacher2018micromagnetics}
J.~Fischbacher, A.~Kovacs, M.~Gusenbauer, H.~Oezelt, L.~Exl, S.~Bance, and
  T.~Schrefl.
\newblock Micromagnetics of rare-earth efficient permanent magnets.
\newblock {\em J. Phys. D: Appl. Phys.}, 51(19):193002, Apr. 2018.
\newblock \url{https://doi.org/10.1088/1361-6463/aab7d1}.

\bibitem{fisher2018model}
A.~Fisher, C.~Rudin, and F.~Dominici.
\newblock Model class reliance: {Variable} importance measures for any machine
  learning model class, from the" rashomon" perspective.
\newblock {\em arXiv preprint arXiv:1801.01489}, 2018.

\bibitem{geron2017hands}
A.~G{\'e}ron.
\newblock {\em Hands-on machine learning with Scikit-Learn and {TensorFlow:}
  {Concepts,} tools, and techniques to build intelligent systems}.
\newblock " O'Reilly Media, Inc.", 2017.

\bibitem{gronefeld1989calculation}
M.~Gr{\"o}nefeld and H.~Kronm{\"u}ller.
\newblock Calculation of strayfields near grain edges in permanent magnet
  material.
\newblock {\em J. Magn. Magn. Mater.}, 80(2-3):223--228, Aug. 1989.
\newblock \url{https://doi.org/10.1016/0304-8853(89)90122-4}.

\bibitem{guptasarma1998new}
D.~Guptasarma and B.~Singh.
\newblock New scheme for computing the magnetic field resulting from a
  uniformly magnetized arbitrary polyhedron.
\newblock {\em GEOPHYSICS}, 64(1):70--74, Jan. 1999.
\newblock \url{https://doi.org/10.1190/1.1444531}.

\bibitem{huber20163d}
C.~Huber, C.~Abert, F.~Bruckner, M.~Groenefeld, O.~Muthsam, S.~Schuschnigg,
  K.~Sirak, R.~Thanhoffer, I.~Teliban, C.~Vogler, R.~Windl, and D.~Suess.
\newblock {3D} print of polymer bonded rare-earth magnets, and {3D} magnetic
  field scanning with an end-user {3D} printer.
\newblock {\em Appl. Phys. Lett.}, 109(16):162401, Oct. 2016.
\newblock \url{https://doi.org/10.1063/1.4964856}.

\bibitem{kronmuller2007general}
H.~Kronm{\"u}ller.
\newblock General micromagnetic theory.
\newblock {\em Handbook of Magnetism and Advanced Magnetic Materials}, 2007.

\bibitem{kronmuller1987angular}
H.~Kronm{\"u}ller, K.-D. Durst, and G.~Martinek.
\newblock Angular dependence of the coercive field in sintered {Fe77Nd15B8}
  magnets.
\newblock {\em J. Magn. Magn. Mater.}, 69(2):149--157, Oct. 1987.
\newblock \url{https://doi.org/10.1016/0304-8853(87)90111-9}.

\bibitem{li2017additive}
L.~Li, B.~Post, V.~Kunc, A.~M. Elliott, and M.~P. Paranthaman.
\newblock Additive manufacturing of near-net-shape bonded magnets: {Prospects}
  and challenges.
\newblock {\em Scr. Mater.}, 135:100--104, July 2017.
\newblock \url{https://doi.org/10.1016/j.scriptamat.2016.12.035}.

\bibitem{mangal2018applied2}
A.~Mangal and E.~A. Holm.
\newblock Applied machine learning to predict stress hotspots i: {Face}
  centered cubic materials.
\newblock {\em Int. J. Plast.}, July 2018.
\newblock \url{https://doi.org/10.1016/j.ijplas.2018.07.013}.

\bibitem{mangal2018comparative}
A.~Mangal and E.~A. Holm.
\newblock A comparative study of feature selection methods for stress hotspot
  classification in materials.
\newblock {\em Integr Mater Manuf Innov}, pages 1--9, June 2018.
\newblock \url{https://doi.org/10.1007/s40192-018-0109-8}.

\bibitem{molnar2018}
C.~Molnar.
\newblock {\em Interpretable Machine Learning}.
\newblock https://christophm.github.io/interpretable-ml-book/, 2018.
\newblock \url {https://christophm.github.io/interpretable-ml-book/}.

\bibitem{murakami2014magnetism}
Y.~Murakami, T.~Tanigaki, T.~Sasaki, Y.~Takeno, H.~Park, T.~Matsuda, T.~Ohkubo,
  K.~Hono, and D.~Shindo.
\newblock Magnetism of ultrathin intergranular boundary regions in
  {Nd}--{Fe}--b permanent magnets.
\newblock {\em Acta Mater.}, 71:370--379, June 2014.
\newblock \url{https://doi.org/10.1016/j.actamat.2014.03.013}.

\bibitem{oikawa2016large}
T.~Oikawa, H.~Yokota, T.~Ohkubo, and K.~Hono.
\newblock Large-scale micromagnetic simulation of {Nd}-{Fe}-b sintered magnets
  with {Dy}-rich shell structures.
\newblock {\em AIP Adv.}, 6(5):056006, May 2016.
\newblock \url{https://doi.org/10.1063/1.4943058}.

\bibitem{quey2011large}
R.~Quey, P.~Dawson, and F.~Barbe.
\newblock Large-scale {3D} random polycrystals for the finite element method:
  {Generation,} meshing and remeshing.
\newblock {\em Comput. Methods Appl. Mech. Eng.}, 200(17-20):1729--1745, Apr.
  2011.
\newblock \url{https://doi.org/10.1016/j.cma.2011.01.002}.

\bibitem{quey2018optimal}
R.~Quey and L.~Renversade.
\newblock Optimal polyhedral description of {3D} polycrystals: {Method} and
  application to statistical and synchrotron x-ray diffraction data.
\newblock {\em Comput. Methods Appl. Mech. Eng.}, 330:308--333, Mar. 2018.
\newblock \url{https://doi.org/10.1016/j.cma.2017.10.029}.

\bibitem{rave1998corners}
W.~Rave, K.~Ramstöck, and A.~Hubert.
\newblock Corners and nucleation in micromagnetics.
\newblock {\em J. Magn. Magn. Mater.}, 183(3):329--333, Mar. 1998.
\newblock \url{https://doi.org/10.1016/s0304-8853(97)01086-x}.

\bibitem{ribeiro2016should}
M.~T. Ribeiro, S.~Singh, and C.~Guestrin.
\newblock Why should i trust you?: {Explaining} the predictions of any
  classifier.
\newblock In {\em Proceedings of the 22nd ACM SIGKDD International Conference
  on Knowledge Discovery and Data Mining}, pages 1135--1144. ACM, 2016.
\newblock \url{https://doi.org/10.1145/2939672.2939778}.

\bibitem{lee1988thirteen}
J.~L. Rodgers and W.~A. Nicewander.
\newblock Thirteen ways to look at the correlation coefficient.
\newblock {\em The American Statistician}, 42(1):59, Feb. 1988.
\newblock \url{https://doi.org/10.2307/2685263}.

\bibitem{sarkar2018practical}
D.~Sarkar, R.~Bali, and T.~Sharma.
\newblock {\em Practical Machine Learning with Python}.
\newblock Apress, 2018.
\newblock \url{https://doi.org/10.1007/978-1-4842-3207-1}.

\bibitem{sasaki2016formation}
T.~Sasaki, T.~Ohkubo, Y.~Takada, T.~Sato, A.~Kato, Y.~Kaneko, and K.~Hono.
\newblock Formation of non-ferromagnetic grain boundary phase in a {Ga}-doped
  {Nd}-rich {Nd}--{Fe}--b sintered magnet.
\newblock {\em Scr. Mater.}, 113:218--221, Mar. 2016.
\newblock \url{https://doi.org/10.1016/j.scriptamat.2015.10.042}.

\bibitem{schrefl1992numerical}
T.~Schrefl and J.~Fidler.
\newblock Numerical simulation of magnetization reversal in hard magnetic
  materials using a finite element method.
\newblock {\em J. Magn. Magn. Mater.}, 111(1-2):105--114, June 1992.
\newblock \url{https://doi.org/10.1016/0304-8853(92)91063-y}.

\bibitem{sepehri2013mechanism}
H.~Sepehri-Amin, T.~Ohkubo, and K.~Hono.
\newblock The mechanism of coercivity enhancement by the grain boundary
  diffusion process of {Nd}--{Fe}--b sintered magnets.
\newblock {\em Acta Mater.}, 61(6):1982--1990, Apr. 2013.
\newblock \url{https://doi.org/10.1016/j.actamat.2012.12.018}.

\bibitem{skokov2018heavy}
K.~Skokov and O.~Gutfleisch.
\newblock Heavy rare earth free, free rare earth and rare earth free magnets -
  vision and reality.
\newblock {\em Scr. Mater.}, 154:289--294, Sept. 2018.
\newblock \url{https://doi.org/10.1016/j.scriptamat.2018.01.032}.

\bibitem{sodervznik2016high}
M.~Soder{\v{z}}nik, M.~Korent, K.~{\v{Z}}. Soder{\v{z}}nik, M.~Katter,
  K.~{\"U}st{\"u}ner, and S.~Kobe.
\newblock High-coercivity {Nd}-{Fe}-b magnets obtained with the electrophoretic
  deposition of submicron {TbF}3 followed by the grain-boundary diffusion
  process.
\newblock {\em Acta Mater.}, 115:278--284, Aug. 2016.
\newblock \url{https://doi.org/10.1016/j.actamat.2016.06.003}.

\bibitem{stoner1948mechanism}
E.~C. Stoner and E.~P. Wohlfarth.
\newblock A mechanism of magnetic hysteresis in heterogeneous alloys.
\newblock {\em Philosophical Transactions of the Royal Society A: Mathematical,
  Physical and Engineering Sciences}, 240(826):599--642, May 1948.
\newblock \url{https://doi.org/10.1098/rsta.1948.0007}.

\bibitem{thompson2017grain}
M.~P. Thompson, E.~Chang, A.~Foto, J.~G. Citron-Rivera, D.~Haddad, R.~Waldo,
  and F.~E. Pinkerton.
\newblock Grain-boundary-diffused magnets: {The} challenges in obtaining
  reliable and representative {BH} curves for electromagnetic motor design.
\newblock {\em IEEE Electrific. Mag.}, 5(1):19--27, Mar. 2017.
\newblock \url{https://doi.org/10.1109/mele.2016.2644561}.

\bibitem{pedregosa2011scikit}
G.~Varoquaux, L.~Buitinck, G.~Louppe, O.~Grisel, F.~Pedregosa, and A.~Mueller.
\newblock Scikit-learn.
\newblock {\em GetMobile: Mobile Comp. and Comm.}, 19(1):29--33, June 2015.
\newblock \url{https://doi.org/10.1145/2786984.2786995}.

\bibitem{yang2017ree}
Y.~Yang, A.~Walton, R.~Sheridan, K.~Güth, R.~Gauß, O.~Gutfleisch, M.~Buchert,
  B.-M. Steenari, T.~Van~Gerven, P.~T. Jones, and K.~Binnemans.
\newblock {REE} recovery from end-of-life {NdFeB} permanent magnet scrap: {A}
  critical review.
\newblock {\em J. Sustain. Metall.}, 3(1):122--149, Sept. 2016.
\newblock \url{https://doi.org/10.1007/s40831-016-0090-4}.

\bibitem{zickler2017combined}
G.~A. Zickler, J.~Fidler, J.~Bernardi, T.~Schrefl, and A.~Asali.
\newblock A combined {TEM/STEM} and micromagnetic study of the anisotropic
  nature of grain boundaries and coercivity in {Nd}-{Fe}-b magnets.
\newblock {\em Adv. Mater. Sci. Eng.}, 2017:1--12, 2017.
\newblock \url{https://doi.org/10.1155/2017/6412042}.

\end{thebibliography}
\end{document}